\documentstyle[amsmath,amssymb,aps,multicol,epsfig,prb,float]{revtex}

\draft

\def\beginwide{
  \end{multicols} \vspace*{-0.5cm} \noindent
  \rule{3.5in}{.1mm}\rule{.1mm}{5mm} \widetext \medskip }
\def\endwide{
  \hspace*{3.5in}~\rule[-5mm]{.1mm}{5mm}\rule{3.5in}{.1mm}
  \begin{multicols}{2}\narrowtext \vspace*{-1.0cm} \noindent }

\def\beginwideapp{
  \end{multicols} \vspace*{-0.5cm} \noindent
  \rule{3.5in}{.1mm}\rule{.1mm}{5mm} \widetext \medskip 
  \renewcommand{\theequation}{\mbox{\Alph{section}\arabic{equation}}}
  \renewcommand{\thesection}{\mbox{\Alph{section}}}
  }
\def\endwideapp{
  \hspace*{3.5in}~\rule[-5mm]{.1mm}{5mm}\rule{3.5in}{.1mm}
  \begin{multicols}{2}\narrowtext \vspace*{-1.0cm} \noindent}

\newcommand{\1}{{(1)}}
\newcommand{\2}{{(2)}}

\newcommand{\p}{\partial}

\newcommand{\dl}{{\frac d{d \ell}}}
\newcommand{\dls}{{\left. \frac d{d \ell} \right|_{\rm scal}}}
\newcommand{\dli}{{\left. \frac d{d \ell} \right|_{\rm int}}}

\newcommand{\cum}{{\rm c}}

\newcommand{\bk}{{\bf k}}
\newcommand{\bN}{{\bf 0}}
\newcommand{\bx}{{\bf x}}

\newcommand{\cH}{{\cal H}}
\newcommand{\cHe}{{\cal H}_{\rm el}}
\newcommand{\cHp}{{\cal H}_{\rm pin}}
\newcommand{\cO}{{\cal O}}
\newcommand{\cZe}{{\cal Z}_{\rm el}}

\newcommand{\ho}{\textrm{h.o.}}

\newcommand{\grad}{{\bbox \nabla}}

\newcommand{\Rp}{R_+}
\newcommand{\Rpp}{R_{++}}
\newcommand{\Smm}{S_{--}}

\begin{document}

\title{Interface fluctuations in disordered systems: \\
  Universality and non-Gaussian statistics}

\author{Stefan Scheidl and Yusuf Din\c{c}er}

\address{Institut f\"ur Theoretische Physik, Universit\"at zu
  K\"oln, Z\"ulpicher Str. 77, D-50937 K\"oln, Germany}

\date{\today}

\maketitle

\begin{abstract}
  We employ a functional renormalization group to study interfaces in
  the presence of a pinning potential in $d=4-\epsilon$ dimensions.
  In contrast to a previous approach [D.S. Fisher, Phys. Rev. Lett.
  {\bf 56}, 1964 (1986)] we use a soft-cutoff scheme.  With the method
  developed here we confirm the value of the roughness exponent $\zeta
  \approx 0.2083 \epsilon$ in order $\epsilon$.  Going beyond previous
  work, we demonstrate that this exponent is universal. In addition,
  we analyze the generation of higher cumulants in the disorder
  distribution and the role of temperature as a dangerously irrelevant
  variable.
\end{abstract}

\pacs{PACS numbers: 46.65.+g, 11.10.Gh, 75.10.Nr}

\begin{multicols}{2} \narrowtext


\section{Introduction}

Quenched disorder plays a crucial role in a huge variety of physical
systems.  One of the most prominent examples for such systems is a
domain wall in Ising-like systems.  Such interfaces can couple to
various types of disorder such as a random pinning potential which can
be provided by impurities in the system.  For interfaces with internal
dimensions $d \leq 4$, such disorder is known to radically modify the
structure of the interface and also its dynamics.  The structure of
such systems typically is self-affine, resembling pure systems at
criticality.  In contrast to the statics, the dynamics is very
different from a critical one, since it lacks the characteristic
scaling relations.  It is dominated by high energy barriers which lead
to an exponentially slow, glassy dynamics.  This behavior was observed
experimentally by position-space imaging techniques \cite{Lemerle+98}
and it should also be possible to study such systems by means of
scattering techniques.\cite{Salditt+95}

The theoretical understanding of such systems has made substantial
progress during the last 15 years.  Although the self-affinity of the
structure suggests the use of renormalization-group (RG) techniques to
calculate the characteristic exponents (such as the roughness
exponent), the analysis intricate since the flow of an infinite number
of relevant parameters has to be studied.  This can be achieved in the
framework of a {\em functional} renormalization group which was
employed by D.S.  Fisher\cite{FisherDS86:dr} on the basis of the {\em
  hard-cutoff renormalization group} (HCRG) scheme of Wegner and
Houghton.\cite{Wegner+73} This study was the prototype for subsequent
generalizations to a variety of different physical systems with a
higher number of components of the displacement field\cite{Balents+93}
or to periodic systems\cite{Giamarchi+94,Emig+99}.  The latter class
includes systems such as charge-density waves, Wigner crystals, and
vortex lines in type-II superconductors.  In all these systems
disorder plays a qualitatively similar role.

However, this HCRG is known to suffer from pathologies related to the
sharp cutoff as noted already in Ref. \CITE{FisherDS86:dr}.  This
cutoff procedure leads to a long-ranged and oscillatory behavior of
the field correlations in real space, which leads to the generation of
highly {\em nonlocal} interactions on a coarse-grained level.  It
requires particular care to interpret these interactions as the
renormalization of local quantities such as the interface stiffness.
These pathologies emerge not only in the context of disordered systems
but also in pure systems, for example in the sine-Gordon model which
describes the roughening transition of crystal surfaces.  In the
latter context, it was pointed out by Nozi\`eres and
Gallet\cite{Nozieres+87} that these pathologies may affect the
renormalization of the stiffness constant and ultimately the scaling
behavior at criticality.

For this reason it is of principal interest to reinvestigate the
prototype model for disordered systems in the framework of a {\em
  soft-cutoff renormalization group} (SCRG).  In this scheme
fluctuations are regularized on small length scales by a smooth cutoff
function.  There exist several variants of SCRGs (for recent review
articles on RG schemes, see e.g. Refs. \CITE{Chang+92,Berges+00}).
The scheme of Wilson and Kogut\cite{Wilson+74} is a SCRG.
Unfortunately, it is very clumsy to work with since already the free
interface is described by a rather complicated Gaussian fixed point.
We use the scheme of Polchinski\cite{Polchinski84}, which we find
convenient for our purposes.

Starting from this SCRG method we confirm the value of the roughness
exponent found in the HCRG scheme.\cite{FisherDS86:dr} In addition, we
explicitly demonstrate its universality, i.e. its independence from
the cutoff function.  In this respect our calculation parallels the
demonstration of universality for the bulk exponent $\eta$ of the pure
Ising model.\cite{Golner+75,Shukla+75,Rudnick75}

In most theoretical analyses disorder is assumed to be Gaussian
distributed, i.e. that higher cumulants of the disorder distribution
are negligible.  Within the present scheme higher cumulants play a
central role.  Although in general these cumulants are non-universal,
we determine their functional form at the RG fixed-point point since
they are physically meaningful for energy fluctuations on large scales
and since they will be needed for the analysis of the model to
subleading order in $\epsilon$.

In the next section \ref{sec.mod} we specify the model.  In order to
keep this paper self-contained we include a brief scaling analysis in
Sec. \ref{sec.scal}.  The application of the Polchinski SCRG scheme to
the present model is discussed in Sec. \ref{sec.scheme}.  The
renormalization flow equations are derived and evaluated in Sec.
\ref{sec.order.2} and our results are discussed in Sec.
\ref{sec.disc}.


\section{Model}
\label{sec.mod}

Our analysis is based on the model Hamiltonian 
\begin{equation}
  \cH=\int d^d x \left\{ \frac \gamma 2 (\grad \phi)^2
    + V(\phi(\bx),\bx) \right\},
  \label{orig.model}
\end{equation}
which is composed of an elastic energy and a pinning energy.  $\gamma$
is the elastic stiffness constant of the interface.  $V$ represents a
quenched random potential acting on the interface.  In the simplest
case the disorder potential $V$ is Gaussian distributed with zero
average and short-ranged correlations in its dependence on $\bx$ and
$\phi$,
\begin{equation}
  \overline{V(\phi_1, \bx_1)V(\phi_2, \bx_2)}  =
  R(\phi_1 - \phi_2) \delta^{(d)}(\bx_1 -\bx_2).
  \label{corr.R}
\end{equation}
In principle one could allow for a more general form of the correlator
(\ref{corr.R}), where the dependence on $\bx$ and $\phi$ does not
factorize and which has a finite correlation length also in $\bx$
directions.  However, in a coarse-grained description this correlation
length shrinks to zero and one ends up with a dependence of the form
(\ref{corr.R}).  Therefore this correlation length can be taken as
zero right away without a modification of the large-scale properties
of the system.  On the contrary, the $\phi$ dependence of the
correlator has to be described by the function $R(\phi)$ with a {\em
  finite} width.  Otherwise the typical force density $-\frac
{\partial}{\partial \phi} V(\phi,\bx)$ would be infinite and the
problem would be ill defined.

We subsequently focus on disorder of the random-potential type, where
$R(\phi)$ decays for large values of $\phi$.  Thus, we exclude systems
with random-field disorder, for which many large-scale features can be
derived already from Imry-Ma type scaling
arguments.\cite{Grinstein+82} We further specialize to {\em short
  ranged} random potentials where $R(\phi)$ decays faster than any
power of $\phi$.  In addition, we assume a statistical reflection
symmetry $\phi \to - \phi$ such that the correlation function is even,
$R(-\phi)=R(\phi)$. 

For the analysis of the model it is convenient to apply the standard
replica trick \cite{Edwards+75} in order to anticipate the disorder
average and to restore translation symmetry. After replicating the
system $n$ times and averaging over disorder, one obtains the
Hamiltonian
\begin{mathletters}
  \label{model.n}
  \begin{eqnarray}
    \cH &=& \cHe+ \cHp ,
    \\
    \cHe &=&  
    \frac{\gamma}{2}  \sum_{\alpha=1}^n \int d^dx \ (\grad \phi^\alpha)^2,
    \\
    \cHp &=& -\frac 1{2 T} \sum_{\alpha, \beta=1}^n 
    \int d^dx \ R(\phi^\alpha(\bx)- \phi^\beta(\bx)),
    \label{Hi_initial}
  \end{eqnarray}
\end{mathletters}
which we have decomposed into an ``elastic'' part and a ``pinning''
part.  In this formulation, the quenched disorder is represented by an
effective interaction between different replicas.  $T$ denotes the
temperature of the system.

Since we have taken the correlation length in $\bx$ direction as zero,
the interaction is local in $\bx$ and the interaction energy is
invariant under an arbitrary replica-independent tilt $\phi^\alpha
(\bx) \to \phi^\alpha (\bx) + \delta \phi(\bx)$ of the interface.  This
symmetry will play an important role below.

As usual, the partition function can be written as a functional
integral
\begin{equation}
  {\cal Z} = {\rm Tr}_\phi \ e^{-\frac{1}{T} \cH[\phi]}
  \label{intro_Z}
\end{equation}
with ${\rm Tr}_\phi=\prod_{\bk} \prod_\alpha \int d \phi^\alpha (\bk)$
as integral over the Fourier modes $\phi(\bk)= \int d^dx \ e^{-i \bk
  \cdot \bx} \phi(\bx)$.  This model has to be regularized at large
momenta by a momentum cutoff $\Lambda$ related to a microscopic length
$a=2 \pi /\Lambda$.  Our particular choice for the regularization will
be discussed in Sec.  \ref{sec.regular}.


\section{Scaling analysis}
\label{sec.scal}

Before we study this model by means of the renormalization group
technique it is instructive to perform a scaling analysis.  Although a
lucid presentation of this analysis can be found in Ref.
\CITE{Balents+93}, we give a brief summary thereof in order to make
this article self-contained.

To describe the shape fluctuations of the interface we are primarily
interested in the pair correlation
\begin{eqnarray}
  \overline{\langle \phi(\bk) \phi(\bk')\rangle} = \lim_{n \to 0}  
  \langle \phi^\alpha(\bk) \phi^\alpha(\bk')\rangle
  =: C(\bk) \delta(\bk+\bk')
  \label{intro.C}
\end{eqnarray}
(here $\alpha$ is not summed over) after averaging over thermal
fluctuations (denoted by $\langle \cdots \rangle$) and over the
disorder distribution (denoted by $\overline {\cdots}$).  Since the
latter average has been anticipated in the replicated system
(\ref{model.n}) it no longer appears in the central term in Eq.
(\ref{intro.C}).

On large length scales $k \ll \Lambda$ the displacement correlation is
found to be self-affine,
\begin{eqnarray}
  C(\bk) \propto  \frac T \gamma k^{-d -2\zeta}
\end{eqnarray}
with a roughness exponent $\zeta$.  For $\zeta >0$, the interface is
{\em rough} and the relative displacement 
\begin{eqnarray}
  w(\bx-\bx') &:=& \overline{\langle [\phi(\bx)-\phi(\bx')]^2\rangle} 
  \nonumber \\ 
  &=&  \lim_{n \to 0}
  \langle [\phi^\alpha(\bx)-\phi^\alpha(\bx')]^2\rangle 
  \label{intro.w}
\end{eqnarray}
increases like 
\begin{eqnarray}
  w(\bx) \sim  x^{2  \zeta},
  \label{intro.zeta.2}
\end{eqnarray}
on large scales $x \gg a$. We always assume $\zeta<1$ since otherwise
the model breaks down since higher powers of $\grad \phi$ become
relevant in the elastic energy.  For $\zeta<0$, the interface is {\em
  flat} and $w(\bx)$ converges to a finite value $w_\infty$ for $\bx
\to \infty$ and
\begin{eqnarray}
  w_\infty-w(\bx) \sim  x^{2  \zeta}.
\end{eqnarray}

In order to examine the relevance of disorder we analyze the
properties of the model under a rescaling of lengths, field, and
temperature:
\begin{mathletters}
  \label{rescal}
  \begin{eqnarray}
    \bx &=& e^\ell \ \bx_\ell,
    \\
    \phi^\alpha(\bx) &=& e^{\zeta \ell} \  \phi_\ell^\alpha(\bx_\ell),
    \\
    T &=& e^{\theta \ell} \ T_\ell.
    \label{intro.T_l}
  \end{eqnarray}
\end{mathletters}
Here we have introduced the scaling parameter $\ell$ -- which can be
viewed as logarithmic length scale -- and the energy scaling exponent
$\theta$.  The role of $\theta$, which is absent in usual critical
phenomena, will become clear soon.  The scaling hypothesis requires
the statistical weights and therefore ${\cal H}/T$ to be invariant
under rescaling.  The elastic energy remains invariant only if the
stiffness is rescaled according to $\gamma_\ell =
e^{(d-2+2\zeta-\theta) \ell} \ \gamma$, which reads in differential
form
\begin{eqnarray}
  \dls  \gamma_\ell  = (d-2+2\zeta-\theta) \gamma_\ell .
\end{eqnarray}
Thus the stiffness is scale invariant for 
\begin{eqnarray}
  \theta=2 \zeta +d-2.
  \label{theta.val}
\end{eqnarray}
In the absence of disorder one can achieve the scale invariance of
both $\gamma$ and $T$ with the choice $\theta=0$ and
$\zeta=\zeta_{\rm th}:=\frac{2-d}2$.  

Since in the absence of disorder the interface is flat for $d>2$ one
may analyze the {\em relevance of disorder} (i.e.  of $\cHp$)
performing a Taylor expansion of $R(\phi)=\sum_{m\geq 0}
\frac{R_{2m}}{(2m)!}  \phi^{2m}$, assuming analyticity of the
(unrenormalized) correlator $R$ (compare Ref.  \CITE{Balents+93}). A
scale invariance of the statistical weights would require
\begin{eqnarray}
  \dls  R_{2m,\ell} = (d+2m\zeta-2\theta) R_{2m,\ell}.
\end{eqnarray}
If we insert $\zeta=\zeta_{\rm th}$ and $\theta=\theta_{\rm th}=0$
into this equation, we find $\dls R_{2m,\ell} = (d+m(2-d))
R_{2m,\ell}$.  In particular, $\dls R_{2,\ell} = 2 R_{2,\ell}$ and we
conclude that disorder is relevant.  The couplings $R_{2m,\ell}$ with
$m>1$ are less relevant for $d>2$ under rescaling with the {\em
  thermal} exponents.

The scale invariance of both $\cHe/T$ and $\cHp/T$ can be achieved
only for $\theta \neq 0$.  Requiring $\dls \gamma_\ell=0$ and $\dls
R_{2,\ell} =0$ one finds $\theta=\theta_{\rm rf}=:2$ and
$\zeta=\zeta_{\rm rf}:=\frac{4-d}2$, which implies the roughness of
the interface in $d<4$.  This ``random-force'' value of the roughness
exponent is the value one finds in a perturbative treatment of
disorder,\cite{Larkin70:t,Efetov+77:t} where the dependence of the
pinning force on $\phi$ is neglected an which is represented by the
Hamiltonian
\begin{eqnarray}
  \cH_{\rm rf}= \sum_{\alpha \beta} \int d^dx \left\{ \frac \gamma 2
    \delta_{\alpha \beta} ({\bbox \nabla} \phi^\alpha)^2 
    - \frac {R_2}{4T}(\phi^\alpha -\phi^\beta)^2 \right\}
  \label{H_rf}
\end{eqnarray}
which is the contribution to (\ref{model.n}) bilinear in the
displacement.

If we reexamine the relevance of the couplings $R_{2m}$ with the
``random-force'' exponents, we find $\dls R_{2m,\ell} = (4-d)(m-1)
R_{2m,\ell}$.  Thus the couplings $R_{2m}$ with $m>1$ are irrelevant
only in $d>4$. Their relevance in $d<4$ indicates that the roughness
cannot be obtained from perturbation theory and that the correct value
of the roughness exponent most likely is not given by $\zeta_{\rm
  rf}$.

In order to obtain the roughness exponent in $d \leq 4$, a
renormalization group (RG) calculation is required.  This RG has to be
a {\em functional} RG since {\em all} terms in the Taylor series of
$R(\phi)$ are relevant, as $\zeta>0$ and the relevance of coefficients
$R_{2m}$ increases with increasing $m$ for {\em any} value of
$\zeta>0$.

Before we start such a RG calculation, it is worthwhile to mention
that a good estimate of the scaling exponent in $d\leq 4$ can be
obtained from the Flory argument.\cite{Imry+75,Kardar87,Nattermann87}
In this argument one assumes that for a rough interface the
short-ranged disorder correlator can be approximated by $R(\phi)
\simeq R^* \delta(\phi)$ with a weight $R^* := \int d\phi \ R(\phi)$.
Then $\dls R^*_\ell = (d-\zeta - 2 \theta) R^*_\ell$.  The requirement
that $\dls \gamma_\ell=0$ and $\dls R^*_\ell =0$ then leads to the
Flory value of the roughness exponent $\zeta=\zeta_{\rm
  F}:=\frac{4-d}5$.


\section{Renormalization Group}
\label{sec.scheme} 

In $d>4$ the couplings $R_{2m}$ are irrelevant for $m>1$ and the
large-scale properties of the model are governed by the Gaussian
random-force model (\ref{H_rf}). In analogy to usual critical
phenomena (see, e.g. Refs. \CITE{Wilson+74,Wegner76}) we assume that
in $d=4-\epsilon<4$ dimensions the large-scale fluctuations are
described by a fixed point which is close to this Gaussian fixed point
for small $\epsilon$.  However, unlike for usual critical phenomena,
this will be a {\em zero-temperature fixed point}:\cite{FisherDS86:dr}
According to Eq.  (\ref{intro.T_l}), which is equivalent to
\begin{eqnarray}
   \dls T_\ell = - \theta T_\ell, 
   \label{flow.T}
\end{eqnarray}
the effective scale-dependent temperature $T_\ell$ flows to zero on
large length scales $x \propto e^\ell$ provided disorder increases the
roughness of the interface, i.e. $\zeta > \zeta_{\rm th}$ and
$\theta>0$ according to Eq.  (\ref{theta.val}).

\subsection{Rescaling}

Since we must perform a {\em functional} RG analysis, we first
reformulate the flow of the couplings (such as $\gamma$ and $R_{2m}$
in section \ref{sec.scal}) resulting from the rescaling (\ref{rescal})
in a closed form: \cite{note.scal}
\begin{eqnarray}
  \dls  \cH &=&  \sum_\alpha \int d^dx \ 
  \{- \bx \cdot {\bbox \nabla} \phi^\alpha(\bx) + \zeta \phi^\alpha({\bx}) \} 
  \frac{\delta \cH}{\delta \phi^\alpha({\bx})}
  \nonumber \\
  &&+ \theta T^2 \frac{\partial }{\partial T} \frac{\cH}T .
  \label{gen.scal}
\end{eqnarray}
Here and henceforth we drop the subscript $\ell$ for simplicity of
notation.  In writing $\dls$ we stress that this is only the {\em
  scaling} contribution to the flow.  To obtain the full RG flow, this
contribution has to be combined with a contribution $\dli$ which
arises from {\em integrating out} modes of the field $\phi$.

\subsection{Regularization}
\label{sec.regular}

To calculate this second contribution, Fisher \cite{FisherDS86:dr} and
Balents and Fisher \cite{Balents+93} have used the ``hard cutoff''
scheme of Wegner and Houghton \cite{Wegner+73} in their analysis of
the present problem.  We choose the ``soft cutoff'' scheme of
Polchinski \cite{Polchinski84,Zinn-Justin93} for the reasons outlined
in the introduction.

We regularize the theory by modifying the propagator using Schwinger's
proper time method (see, e.g., Ref. \CITE{Zinn-Justin93}).  To this
end, we rewrite $\cHe$ in Fourier space,
\begin{equation}
  \cHe = \frac{1}{2} (\phi,G^{-1}_\Lambda \phi) :=
  \frac{1}{2} \sum_\alpha \int \frac{d^dk}{(2\pi)^d} \ 
  G^{-1}_\Lambda({\bf k}) |\phi^\alpha({\bf k})|^2 .
\end{equation}
Herein the propagator
\begin{equation}
  G_\Lambda(\bk) := G_\infty(\bk) f(k/\Lambda) 
  \text{ with } G_\infty(\bk)  = 1/(\gamma k^2)
\end{equation}
is regularized by the cutoff function $f$.  In order to suppress
fluctuations on short length scales, $f(z)$ has to vanish for large
$z$, whereas it has to satisfy $f \to 1$ for $z \to 0$ in order not to
modify the properties of the model on large length scales.
$f(k/\Lambda)$ can roughly be interpreted as ``weight of modes'': In
the calculation of the pair correlation (\ref{intro.w}) one may
consider the density of modes in Fourier space as being reduced to a
fraction $f$ with respect to the unregularized model.

We will keep the function $f$ general as long as possible, which is
desirable to verify the independence of the results on the
regularization procedure.  However, we implicitly assume $f$ to be
monotonous in order to have smooth functions and to have no further
intrinsic length scales.  For illustrative purposes we occasionally
choose the specific form
\begin{eqnarray}
  f(z)=e^{-z^2/2}.
  \label{f.Gauss}
\end{eqnarray}
As long as $f^{-1}(k/\Lambda)$ is analytic, a Taylor expansion of
$f^{-1}(k/\Lambda)$ shows that the regularization is achieved by
contributions to the Hamiltonian which are irrelevant on large length
scales since they involve higher orders of $k$.  The HCRG method of
Wegner and Houghton \cite{Wegner+73} can be considered as non-analytic
choice $f(z)=\Theta(1-z)$.

\subsection{Mode integration}

In Polchinski's scheme,\cite{Polchinski84} the mode integration works
as follows (see also Ref. \CITE{Zinn-Justin93}).  One introduces an
additional field $\phi^<$ in the partition sum and defines
$\phi^>:=\phi-\phi^<$.  This can be achieved in such a way that
\begin{eqnarray}
  \label{zustand1}
  {\cal Z} &\propto& {\rm Tr}_{\phi^<} 
  e^{-\frac{1}{2T}(\phi^<,G_<^{-1}\phi^<)}
  \nonumber \\
  &&\times {\rm Tr}_{\phi^>} e^{-\frac{1}{2T}(\phi^>,G_>^{-1}\phi^>)
    -\frac 1T \cHp[\phi^<+\phi^>]}.
\end{eqnarray}
The ``slow modes'' $\phi^<$ are regularized by the propagator $G_< :=
G_{\Lambda_<}$ with an infinitesimally reduced cutoff $\Lambda_< :=
e^{-d\ell} \Lambda$, whereas the ``fast modes'' $\phi^>$ have a
propagator $G_> := G_\Lambda-G_<$.  In Eq. (\ref{zustand1}) --- which
is derived from Eq.  (\ref{intro_Z}) in Appendix \ref{app.int} ---
proportionality factors independent of $\phi$ have been dropped.

In the representation (\ref{zustand1}) the modes $\phi^>$ can be
integrated out.  This can be done {\em exactly} in the limit, where
$G_>$ is small (it is of order $d\ell$) since then also $\phi^>$ is
small (of order $(d\ell)^{1/2}$).  Then an expansion of
$\cHp[\phi^<+\phi^>]$ to second order in $\phi^>$ is sufficient to
establish the differential RG.  For infinitesimal $d \ell$ the
propagator of the fast modes is
\begin{mathletters}
  \label{intro.g}
  \begin{eqnarray}
    G_>(\bk)  &=& g(\bk) d\ell + \cO(d \ell^2),
    \label{G_>}
    \\
    g(\bk) &=& \Lambda \frac{d}{d\Lambda} G_\Lambda(\bk) .
  \end{eqnarray}
\end{mathletters}
Integration over the field $\phi^>$ then yields a flow (see Appendix
\ref{app.int}):
\begin{eqnarray}
  \dli \cH &=& 
  \frac{1}{2} \sum_\alpha
  \int_{12} g_{12} \left[ T 
    \frac{\delta^2 \cHp}{\delta \phi_1^\alpha   \delta \phi_2^\alpha }
    - \frac{\delta \cHp}{\delta \phi_1^\alpha }
    \frac{\delta \cHp}{\delta \phi_2^\alpha }
  \right]  .
  \label{gen.int}
\end{eqnarray}
We introduce the abbreviations $\int_{ijk\dots}:= \int_i \int_j \int_k
\dots$, $\int_i := \int d^dx_i$, $\phi_i := \phi(\bx_i)$ and $g_{ij}
:= g(\bx_i-\bx_j)$ to keep expressions compact.  We call the first
term in Eq. (\ref{gen.int}) the ``contraction'' part of the generator
(since legs of a vertex are contracted in a diagrammatic
representation) and the second term the ``composition'' part (since
new vertices are composed by linking two vertices).

The flow contribution (\ref{gen.int}) has a {\em simpler} form than
the corresponding contribution in the HCRG scheme, since
(\ref{gen.int}) contains terms only up to first order in $g$ (and
second order in the Hamiltonian), whereas the HCRG generator contains
terms of arbitrarily high order.  This difference is due to the fact
that $G_>$ is a bounded function of order $d\ell$, in contrast to the
HCRG scheme, where $G_>$ is singular for momenta at the cutoff, for
which reason the power counting of orders in $d\ell$ breaks down and
also higher powers in $G_>$ contribute to ${\cal O}(d\ell)$. The
simplicity of the generator is one important argument in favor of the
SCRG scheme.

The complete functional RG consists of the combination of integration
over modes (\ref{gen.int}) with rescaling (\ref{gen.scal}),
\begin{eqnarray}
  \dl \cH =  \dls \cH + \dli \cH .
  \label{full.gen}
\end{eqnarray}
Although we have defined the regularization in Fourier space, we
ultimately find it more convenient to evaluate the RG flow in position
space because of the functional structure of $\dli \cH$.

To close this exposition of the RG method we emphasize that the RG
(\ref{full.gen}) is {\em exact}, provided the dependence of the
functional $\cHp[\phi^<+\phi^>]$ on small $\phi^>$ (which is of
order $(d\ell)^{1/2}$because of Eq. (\ref{G_>})) is captured in order
$d\ell$ by a functional Taylor expansion to second order in $\phi^>$.
This {\em analyticity} assumption is common to HCRG and SCRG approaches
and will be reexamined later on after we have found the fixed point
for the disordered interface and the generation of nonanalytic
features.

\subsection{General features}
\label{sec.gen}

We now turn to the evaluation of the RG flow for our model
(\ref{model.n}). Due to the locality of the disorder correlator in
$\bx$ space, the model owns a stochastic symmetry under a tilt
$\phi^\alpha(\bx) \to \phi^\alpha(\bx) + {\bf v} \cdot \bx$ with a
constant vector ${\bf v}$.\cite{Schulz+88,Balents+93} As a consequence
of this symmetry there is no renormalization of the stiffness $\gamma$
arising from the mode integration.  Hence the flow of $\gamma$ arises
only from rescaling,
\begin{eqnarray}
  \dl \gamma = (d-2+2\zeta-\theta) \gamma.
\end{eqnarray}
This implies the validity of (\ref{theta.val}) not only in a scaling
analysis but also at the RG fixed point.

In contrast to $\gamma$, the couplings of the disorder part of the
Hamiltonian, $\cHp$, will not only be rescaled but also renormalized
by the mode integration.  This renormalization results not only in a
modification of the second-order cumulant $R$ of the disorder
distribution but also in the generation of contributions to $\cHp$ of
an extended functional form.  The functional form of the emerging
terms can be recognized from the action of the generator
(\ref{gen.int}) on $\cHp$.  The unrenormalized functional
(\ref{Hi_initial}) evaluates the field in {\em two} different replicas
but only at a {\em single} point $\bx$.  Therefore we call functionals
of this type ``2-replica'' functionals and ``1-point'' functionals.
The action of the first ``contraction'' term in (\ref{gen.int})
results again in a 1-point functional.  However, the ``composition''
term evaluates the field at {\em two different} positions, it is a
2-point and 3-replica term.  Fortunately, at the expense of the more
complicated functional form we get ``smaller'' terms (i.e. of higher
order in $R$). This 2-point term is of second order in $R$.
Successive iterations generate $p$-point and $(p+1)$-replica terms in
order $R^p$.

The extended functional form of the Hamiltonian can be captured 
in the form 
\begin{eqnarray}
  \frac 1T \cHp &=& \frac 1T{\cal Q} -\frac 1{2 T^2} 
  {\cal R} + \frac 1{3! T^3} {\cal S}  - \cdots
  \label{cumulants}
\end{eqnarray}
where ${\cal Q}$, ${\cal R}$, and ${\cal S}$ are 1-, 2-, and 3-replica
terms, which represent cumulants of the disorder distribution.  To
become specific, let us denote the pinning energy of the replicated
system before disorder averaging by ${\cal H}_V=\sum_\alpha \int_\bx
V(\phi^\alpha (\bx),\bx)$.  Then the contributions to the extended
Hamiltonian can be identified as cumulants of ${\cal H}_V$:
\begin{eqnarray}
  {\cal Q}=\langle {\cal H}_V \rangle, \
  {\cal R}=\langle {\cal H}_V^2 \rangle_\cum, \
  {\cal S}=\langle {\cal H}_V^3 \rangle_\cum, \dots
\end{eqnarray}
We will not keep track of ${\cal Q}$ since this is a field-independent
constant because of the stochastic symmetry $\phi(\bx) \to \phi(\bx) +
\delta \phi$ for an arbitrary constant $\delta \phi$.

Anticipating that $R=\cO(\epsilon)$ at the fixed point, it is
sufficient to retain $p$-point and $(p+1)$-replica terms to study the
RG flow in $\cO(\epsilon^p)$.  As we will see in Sec.
\ref{sec.order.2}, it is possible to retain the full functional form
of the Hamiltonian without need for truncations.  To keep track of
this functional form it is more convenient to perform the RG analysis
in position space than in momentum space.


\section{RG flow and fixed point}
\label{sec.order.2}

In this section we analyze the RG flow in order $\epsilon^2$, from
which we obtain the roughness exponent $\zeta$ in order $\epsilon$.
We find agreement with previous HCRG calculations
\cite{FisherDS86:dr,Balents+93} in this exponent.  In addition, we are
able to demonstrate the universality of this exponent in the sense of
its independence on the cutoff function $f$.  We also explicitly
obtain the third-order cumulant ${\cal S}$ of the disorder
distribution.

In the HCRG analysis\cite{FisherDS86:dr,Balents+93} it was not
necessary to keep track of 3-replica terms for the determination of
$\zeta$ in order $\epsilon$.  Although 3-replica terms are generated,
they do not feed back into the equation that determines $\zeta$ (in
order $\epsilon$) and there is no need to keep track of ${\cal S}$.
The situation is different in the SCRG scheme, since ${\cal R}$ is
renormalized only via ${\cal S}$.  However, 3-replica terms {\em are}
generated also in the HCRG scheme.  For the analysis of the
consistency of the $\epsilon$-expansion one has to include also this
term into consideration.

In order to determine $\zeta$ in order $\epsilon$, the it is
sufficient to sufficient to consider the Hamiltonian in a functional
form which can be parameterized by functions $\Rp$, $\Rpp$ and $\Smm$
according to
\begin{mathletters} 
  \begin{eqnarray}
    {\cal R} &=&  \sum_{\alpha \beta} \int_1
    \Rp(\phi_1^{\alpha \beta})
    \nonumber \\ &&  
    +  \sum_{\alpha \beta} \int_{12}
    \Rpp(\phi_1^{\alpha \beta},\phi_2^{\alpha \beta};\bx_{12}) + \cdots,
    \\
    {\cal S} &=&\sum_{\alpha \beta \gamma} \int_{12}
    \Smm(\phi_1^{\alpha \beta},\phi_2^{\alpha \gamma};\bx_{12}) + \cdots,
  \end{eqnarray}
\end{mathletters}
where we rewrite $\Rp(\phi) \equiv R(\phi)$ to display that $R$ is an
even function of $\phi$.  To compress notation we further introduce
$\phi_i^{\alpha \beta} := \phi^{\alpha}(\bx_i) - \phi^{\beta}(\bx_i)$.
We thus keep functional contributions up to 3-replica and 2-point
type.  It is not necessary to keep these functionals in the most
general form.  Instead, the functions obey certain symmetry
properties. $\Rpp$ is an even function of both field arguments, $\Smm$
is an odd function of both field arguments (the detailed symmetry
properties are specified below).  Both functions evaluate the fields
at two points and can therefore depend also on the distance
$\bx_{ij}:=\bx_i-\bx_j$ between these points.

A priori there is an ambiguity in representing a functional (such as
${\cal R}$) in terms of $p$-point kernels (such as the 1-point and
2-point functions $\Rp$ and $\Rpp$). For example, the simultaneous
replacement $\Rp(\phi_1^{\alpha \beta}) \to \Rp(\phi_1^{\alpha \beta})
+ \Delta \Rp(\phi_1^{\alpha \beta})$ and $\Rpp(\phi_1^{\alpha
  \beta},\phi_2^{\alpha \beta};\bx_{12}) \to \Rpp(\phi_1^{\alpha
  \beta},\phi_2^{\alpha \beta};\bx_{12}) - \Delta \Rp(\phi_1^{\alpha
  \beta}) \delta(\bx_{12})$ leaves the functional ${\cal R}$
unchanged.  In this way one could absorb the 1-point kernel into the
2-point kernel (in general, low-point kernels into higher-point
kernels).  In the case of ${\cal S}$ we avoid introducing a 1-point
kernel.  In the case of ${\cal R}$ we extract the 1-point kernel in
order to ensure that the 2-point kernel contributes to ${\cal R}$ only
in order $\epsilon^2$ (see below).  A unique distinction between the
1-point and higher-point kernels in a functional is achieved by
requirement that the integral over higher-point kernels must vanish
for a spatially constant field $\phi^{\alpha \beta}(\bx)=
\phi_0^{\alpha \beta}$, in particular $\int d^dx \ \Rpp(\phi_0^{\alpha
  \beta},\phi_0^{\alpha \beta};\bx)=0$.

\begin{figure}
  \epsfig{file=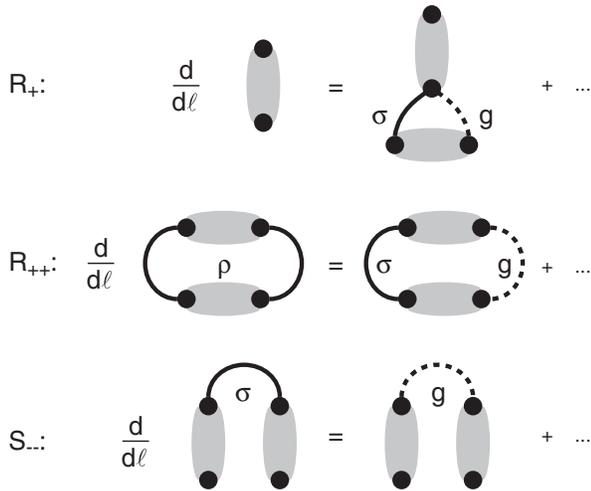,width=0.9 \linewidth}
  \narrowtext
  \caption{Diagrammatic representation of the flow equations 
    (\ref{flow.fun}).  Black points stand for fields
    $\phi^\alpha(\bx)$. A shaded oval ``vertex'' connection between
    two points means that they are evaluated at the same position
    (however, in general their replica index is different) and that
    their difference is the argument of a function.  A line connecting
    two fields represents a kernel function (a broken line represents
    $g$; full lines represent $\sigma$ or $\rho$).  Fields connected
    by a line are confined to the same replica.  Scaling terms and
    topologically equivalent diagrams are not included in this
    illustration.  At the fixed point an oval connection is related to
    a derivative $\Rp^\1$.  The number of lines leaving a vertex
    equals the order of the derivative.  The point order is given by
    the number of vertices.  The replica order is given by the number
    of disconnected groups of fields. Diagrams in which a $g$ line
    would connect points that are already connected via other lines do
    not contribute to the flow equations since they are of higher
    order in temperature.  If both fields of a vertex are linked
    through lines and an odd number of lines leaves the vertex then
    the diagram vanishes for symmetry reasons.}
  \label{fig.diag}
\end{figure}

The flow equation (\ref{full.gen}) for the Hamiltonian can be
represented as flow equation of the parameter functions (for a
diagrammatic representation, see Fig.\ref{fig.diag})
\beginwide
\begin{mathletters}
\label{flow.fun}
\begin{eqnarray}
  \dl \Rp(\phi)&=&\{d-2 \theta + \zeta \phi \p \} \Rp(\phi) +
  \frac23 \int_2 \p_1\p_2 \Smm(\phi_1,0;\bx_{12}) g_{12} 
  \nonumber \\
  & & 
  - \frac13 \int_2 \p_1 \p_2 \Smm(\phi_1,\phi_1;\bx_{12}) g_{12} + \cdots,
  \label{flow.Rp.o2}
  \\
  \dl \Rpp(\phi_1,\phi_2;\bx_{12})&=&
  \{2d-2 \theta + \zeta \phi_i \p_i + \bx_i \cdot \grad_i\}
  \Rpp(\phi_1,\phi_2;\bx_{12}) - 
  \frac{1}{3} \p_1 \p_2 \Smm(\phi_1,\phi_2;\bx_{12}) g_{12} 
  \nonumber \\
  & &
  + \frac13 \delta_{12} \int_3 \p_1 \p_2
  \Smm(\phi_1,\phi_1;\bx_{13}) g_{13} + \cdots,
  \label{flow.Rpp.o2}
  \\
  \dl \Smm(\phi_1,\phi_2;\bx_{12})&=&
  \{2d-3 \theta + \zeta \phi_i \p_i + \bx_i \cdot \grad_i\}
  \Smm(\phi_1,\phi_2;\bx_{12}) 
  - 3 \p \Rp(\phi_1) \p \Rp(\phi_2) g_{12} + \cdots. 
  \label{flow.Smm}
\end{eqnarray}
\end{mathletters}
\endwide
${\bbox \nabla}_{i}:= \frac \p {\p \bx_i}$ is a partial derivative
that acts only on the explicit position arguments $\bx_{ij}$ but not
on the arguments of the fields.  In $\bx_i \cdot \grad_i$ summation
over $i$ is assumed implicitly ($1 \leq i \leq p$ in $p$-point
kernels).  Further on, $\p_i$ denotes the partial derivative acting on
the the $i$th field argument of the kernel function (which is not
necessarily $\phi_i$; if there is only one argument we drop the
subscript, $\p := \p_1$).  

In the flow equations (\ref{flow.fun}) we neglect all terms
proportional to temperature since $T$ flows to zero according to Eq.
(\ref{flow.T}).  One immediately recognizes that even if there is only
a function $\Rp$ in the unrenormalized model, the kernel $\Smm$ is
generated from the composition of two $\Rp$.  The kernel $\Smm$ then
feeds back into the flow equations for $\Rp$ and generates $\Rpp$.

Since $\Rp(\phi)$ and $g(\bx)$ are even functions, $\Smm$ has the
symmetry properties
\begin{eqnarray}
  \Smm(\phi_1,\phi_2;\bx)=\Smm(\phi_1,\phi_2;-\bx)=
  \Smm(\phi_2,\phi_1;\bx)
  \nonumber \\ 
  =-\Smm(-\phi_1,\phi_2;\bx)=-\Smm(\phi_1,-\phi_2;\bx).
  \label{symm.So.2}
\end{eqnarray}
These symmetries immediately imply
\begin{eqnarray}
  \Rpp(\phi_1,\phi_2;\bx)=\Rpp(\phi_1,\phi_2;-\bx)=
  \Rpp(\phi_2,\phi_1;\bx)
  \nonumber \\ 
  = \Rpp(-\phi_1,\phi_2;\bx)=\Rpp(\phi_1,-\phi_2;\bx).
\label{symm.R.2}
\end{eqnarray}

The last terms in Eq.  (\ref{flow.Rp.o2}) and Eq.  (\ref{flow.Rpp.o2})
are an ``insertion of zero'' since their contributions to ${\cal R}$
cancel exactly.  They represent the shift of a 1-point term from the
flow of $\Rpp(\phi_1,\phi_2;\bx_{12})$ to the flow of $\Rp(\phi_1)$ in
order to achieve $\int_\bx \Rpp(\phi_0,\phi_0;\bx) = 0$ according to
the requirement specified above.  This requirement has to be imposed
on $\Rpp$ (unlike $\Smm$) in order to ensure that an $\epsilon$
expansion can be performed consistently.

In order to obtain the roughness exponent to lowest order in
$\epsilon$,
\begin{equation}
  \zeta = \epsilon \zeta^\1 + \ho
  \label{zeta.1}
\end{equation}
(``$\ho$'' stands for higher orders in $\epsilon$) we must determine
the fixed-point Hamiltonian up to order $\epsilon^2$,
\begin{eqnarray}
  \cHp = \epsilon \cHp^\1 + \epsilon^2 \cHp^\2 + \ho.
\end{eqnarray}
To lowest order there is only a 2-replica 1-point contribution
\begin{eqnarray}
  \cHp^\1 = - \frac 1{2T} \sum_{\alpha \beta} \int_1
  \Rp^\1(\phi_1^{\alpha \beta}).
  \label{H.1}
\end{eqnarray}
In order $\epsilon^2$ the RG flow generates 2-point terms which are of
2-replica or 3-replica type:
\begin{eqnarray}
  \cHp^\2 &=& - \frac 1{2T} \sum_{\alpha \beta} \int_1
  \Rp^\2(\phi_1^{\alpha \beta})
  \nonumber \\
  &&- \frac 1{2!T} \sum_{\alpha \beta} \int_{12}
  \Rpp^\2(\phi_1^{\alpha \beta},\phi_2^{\alpha \beta};\bx_{12})
  \nonumber \\
  && + \frac 1{3!T^2} \sum_{\alpha \beta \gamma} \int_{12}
  \Smm^\2(\phi_1^{\alpha \beta},\phi_2^{\alpha \gamma};\bx_{12}).
  \label{H.2}
\end{eqnarray}

Inserting the Hamiltonian (\ref{H.1}) and (\ref{H.2}) into equation
(\ref{full.gen}) we obtain the fixed-point equations for the parameter
functions.  We establish these equations and find their solutions in
the order as they are generated by the RG.

The fixed-point condition $\dl \Smm^\2 (\phi_1,\phi_2;\bx_{12})=0$
reads
\begin{eqnarray}
  0&=&
  \{2 + \bx_i \cdot \grad_i\}
  \Smm^\2(\phi_1,\phi_2;\bx_{12})
  \nonumber \\ &&
  - 3 \p \Rp^\1(\phi_1) \p \Rp^\1(\phi_2) g_{12} . 
  \label{fix.Smm}
\end{eqnarray}
It is solved by a function $\Smm^\2 (\phi_1,\phi_2;\bx_{12})$ in which
the dependences on all arguments factorize:
\begin{eqnarray}
  \label{fix.S.2}
  \Smm^\2(\phi_1,\phi_2;\bx_{12}) = 
  - 3 \p \Rp^\1(\phi_1) \p \Rp^\1(\phi_2) \sigma(\bx_{12})  .
\end{eqnarray}
The explicit spatial dependence is carried by the function $\sigma$
which is determined by the differential equation
\begin{eqnarray}
  2 \sigma(\bx) + \bx \cdot {\bbox \nabla} \sigma(\bx) &=& -g(\bx) .
  \label{dgl.sigma}
\end{eqnarray}
The solution of this differential equation (which we require to
preserve the rotation symmetry in $\bx$ of the unrenormalized model)
involves one constant of integration, which can be fixed by the
requirement that the Fourier transform $\sigma(\bk)$ be analytic at
$\bk=\bN$.  This is a common requirement, which is crucial also for
the solution of the $\phi^4$ model, see Refs.
\CITE{Wegner+73,Wilson+74}.  The Fourier transform of Eq.
(\ref{dgl.sigma}) is $(2-\epsilon + \bk \cdot \grad_\bk)
\sigma(\bk)=g(\bk)$ which is solved by
\begin{eqnarray}
  \sigma(\bk)=\int_0^1 dz \ z^{1- \epsilon}  \ g(z \bk) .
  \label{sol.sigma}
\end{eqnarray}
This solution is analytic at $\bk=\bN$ for $d>2$ since $g(\bk)$ is
analytic.  The explicit form of various appearing kernel functions is
given in appendix \ref{app.kern} for the special cutoff function
(\ref{f.Gauss}).  Since $R^\1$ is of order $\epsilon$, $\sigma$ is
needed only in order $\epsilon^0$ to determine $\Smm^\2$ from Eq.
(\ref{fix.S.2}).  In general, $g (\bk=\bN) \neq 0$ and then also
$\sigma(\bk=\bN) \neq 0$. The $\bk=\bN$ contribution to the functional
$\Smm^\2$ could in principle be split off as a 1-point term.  We
refrain from doing so because both the 1-point term and the 2-point
term are of order $\epsilon^2$ and this would unnecessarily increase
the number of terms.

In order $\epsilon^2$ the fixed-point conditions for $\Rp^\2$ and
$\Rpp^\2$ read 
\beginwide
\begin{mathletters}
  \label{fix.R}
  \begin{eqnarray}
    \label{fix.Rp}
    0 
    &=&
    \{1-4 \zeta^\1 +\zeta^\1 \phi_1 \p \} \Rp^\1(\phi_1) 
    +\frac23 \int_2 \p_1\p_2 \Smm^\2(\phi_1,0;\bx_{12}) g_{12} 
    - \frac13 \int_2 \p_1 \p_2 \Smm^\2(\phi_1,\phi_1;\bx_{12}) g_{12}, 
    \\
    \label{fix.Rpp}
    0
    &=& 
    \{4 + \bx_i \cdot \grad_i \}
    \Rpp^\2(\phi_1,\phi_2;\bx_{12})
    - \frac{1}{3} \p_1 \p_2 \Smm^\2(\phi_1,\phi_2;\bx_{12}) g_{12}
    + \frac13 \delta_{12} \int_3 \p_1 \p_2
    \Smm^\2(\phi_1,\phi_1;\bx_{13}) g_{13}.
  \end{eqnarray}
\end{mathletters}
\endwide
Plugging the solution (\ref{fix.S.2}) into the fixed-point equation
(\ref{fix.Rpp}), $\Rpp^\2$ is found in the form
\begin{eqnarray}
  \Rpp^\2(\phi_1,\phi_2;\bx_{12}) = - \p^2 \Rp^\1(\phi_1) \p^2
  \Rp^\1(\phi_2) \rho(\bx_{12}) 
\end{eqnarray}
with a nonlocal kernel $\rho$ satisfying the equation
\begin{eqnarray}
  4 \rho(\bx) + \bx \cdot {\bbox \nabla} \rho(\bx)
  &=&  c(\bx) - c_0 \delta(\bx) , 
  \label{rho.dif}
\end{eqnarray}
where we defined
\begin{mathletters}
  \begin{eqnarray}
    c(\bx) &:=& \sigma(\bx) g(\bx) , 
    \\
    c_0 &:=& \int_\bx c(\bx) = c(\bk=\bN).
    \label{def.c}
  \end{eqnarray}
\end{mathletters}
The differential equation for $\rho$ is solved in Fourier space in
analogy to Eq.  (\ref{dgl.sigma}) imposing the analyticity requirement
at small wave vectors,
\begin{eqnarray}
  \rho(\bk)= \int_0^1 dz \ z^{-1-\epsilon} \ [c_0-c(z \bk)].
  \label{rho.sol}
\end{eqnarray}
Since $g(\bk)$ is analytic, $\sigma(\bk)$ and $c(\bk)$ are analytic
and so is $\rho(\bk)$.  It is important to notice that it is possible
to find a solution which is analytic and {\em finite} for $\epsilon
\to 0$ only due to the presence of the last term in equation
(\ref{fix.Rpp}). In the absence of this term, $c_0$ would be absent in
Eq. (\ref{rho.sol}) and $\rho(\bk=\bN) \propto 1/\epsilon$ would
become singular for $d \to 4^+$.  This would contradict our assumption
that the 2-point kernel $\Rpp$ contributes to $\cH$ only in order
$\epsilon^2$; instead it would contribute to $\cH$ in order $\epsilon$
and generate even more complicated contributions to $\cH$ in order
$\epsilon^2$.

Now we turn to the determination of the fixed-point function $\Rp^\1$
from equation (\ref{fix.Rp}).  Inserting the given solution for
$\Smm^\2$ into this equation we find
\begin{eqnarray}
  0 &=& \{1 - 4 \zeta^\1 + \zeta^\1 \phi \p \} \Rp^\1(\phi) 
  \nonumber \\
  & & + c_0 [\p^2 \Rp^\1(\phi) \p^2 \Rp^\1(\phi)
  - 2 \p^2 \Rp^\1(\phi) \p^2 \Rp^\1(0)] .
\label{fix.R.1}
\end{eqnarray}
Remarkably, the entire information about the cutoff function $f$ (as
well as the surface stiffness $\gamma$) is contained in the constant
$c_0$.  This constant can be eliminated by a rescaling $R \to R/c_0$,
after which our flow equation coincides with the one derived with the
HCRG scheme.\cite{FisherDS86:dr,Balents+93}

\begin{figure}
  \epsfig{file=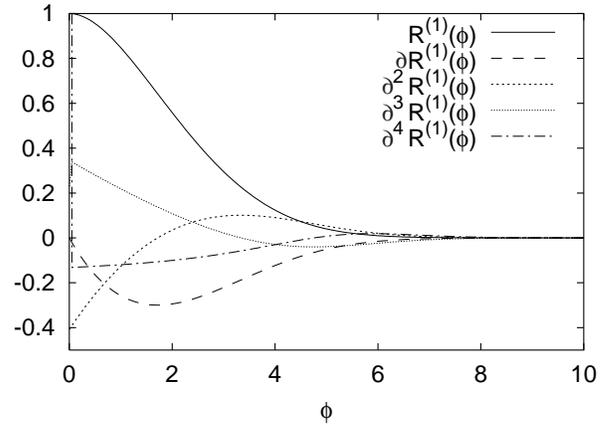,width=0.9 \linewidth}
  \narrowtext
  \caption{Illustration of the fixed-point function $R^\1(\phi)$ and its
    first four derivatives.  Note that $\p^3 R^\1(0^+)>0$. The
    singular contribution to $\p^4 R^\1(\phi)$ is represented
    by the vertical part of the line near $\phi=0$.}
  \label{fig.R1}
\end{figure}

Since the fixed-point equation (\ref{fix.R.1}) is homogeneous, i.e.
invariant under a simultaneous rescaling
\begin{eqnarray}
  \phi \to \widetilde \phi:= b \phi \textrm{ and }
  \Rp(\phi) \to \widetilde \Rp(\widetilde \phi) :=b^4 \Rp(\phi) 
  \label{cont.sol}
\end{eqnarray}
for any real $b$, it has a {\em continuous set} of solutions.
Therefore one can search a solution of Eq.  (\ref{fix.R.1}) (which is
a nonlinear differential equation of second order for $\Rp^\1$)
without loss of generality with the initial condition $\Rp^\1(0)=1$.
The second condition $\p \Rp^\1(0)=0$ follows from the symmetry of
$\Rp$.  The fixed point has to be determined numerically.  For a
trial value of $\zeta^\1$ one integrates Eq. (\ref{fix.R.1}) from
$\phi=0$ to $\phi=\infty$.  The fixed point describing elastic
interfaces in disorder with short-ranged correlations is determined
from the condition that $\Rp^\1(\phi)$ decays monotonously and faster
than any power.  We find
\begin{equation}
  \zeta^\1=0.208298(1)  
\label{zeta1}
\end{equation}
and the fixed-point function and its derivatives are illustrated in
Fig. \ref{fig.R1}.  As discussed in Ref. \CITE{FisherDS86:dr}, from
this flow equation (\ref{fix.R.1}) one actually can determine not only
the fixed point and scaling behavior for short-ranged correlations but
also for long-ranged correlations which we will not further analyze
here.


\section{Results and discussion}
\label{sec.disc}

The RG result (\ref{zeta.1}) with (\ref{zeta1}) is close to the Flory
value $\zeta_{\rm F}=\frac \epsilon 5$.  For $d=1$ the value
$\zeta(d=1)=\frac 23$ of the roughness exponent is known
exactly.\cite{Huse+85:exp} In $d=2$ and $d=3$ numerical
calculations\cite{Middleton95} resulted in the values
$\zeta(d=2)=0.41(1)$ and $\zeta(d=3)=0.22(1)$.  Although it is
questionable whether the corresponding values of $\epsilon=3$ are
small enough to justify the neglection of higher order terms, the
lowest order estimate (\ref{zeta.1}) yields values $\zeta(d=1) \approx
0.6249$, $\zeta(d=2) \approx 0.4166$, and $\zeta(d=3) \approx 0.2083$,
which are surprisingly close to the exact or numerical values.

Since $f$ enters the fixed-point equation only via $c_0$ which can be
eliminated by rescaling, we have explicitly shown {\em universality},
i.e. that $\zeta^\1$ is independent of the cutoff function $f$.
However, the fixed-point Hamiltonian does depend on this function.  In
particular, also the disorder cumulants are nonuniversal.  We wish to
stress again that in our analysis the {\em nonlocality} has been
treated exactly, whereas Fisher\cite{FisherDS86:dr} and Balents and
Fisher\cite{Balents+93} have truncated the flow to spatially constant
fields which is sufficient to determine $\zeta^\1$.  The nonlocality
is important for the evaluation of the RG flow in higher orders of
$\epsilon$.\cite{Dincer99,Scheidl00:un} However, only $\Smm^\2$ feeds
back into the flow equation for $\Rp^\2$ from which $\zeta^\1$ is
determined.  In contrast, $\Rpp^\2$ with the proper definition of
$\rho$ [where $\rho(\bk=\bN)=0$ ensures that $\Rpp$ is of order
$\epsilon^2$] does not feed back.  Nevertheless, it is important to
know $\Rpp^\2$ in order to determine $\zeta$ in higher orders of
$\epsilon$.\cite{Dincer99,Scheidl00:un}

The most distinctive feature of the fixed-point solution is the {\em
  nonanalyticity} of $\Rp^\1$, as pointed out by
Fisher\cite{FisherDS86:dr} and Balents and Fisher\cite{Balents+93} and
which is implied by the fixed-point equation (\ref{fix.R.1}) as
follows.  According to Eq. (\ref{corr.R}) a physically meaningful
solution must satisfy $\Rp^\1(0)>0$ and $\Rp^\1(\phi)$ can be expected
to decay monotonously decaying for increasing $\phi$.  Searching a
solution which is as smooth as possible, we may assume $\p
\Rp^\1(\phi)$ to be continuous and to vanish for $\phi=0$. Evaluating
Eq.  (\ref{fix.R.1}) at $\phi=0$ one then finds $\p^2 \Rp^\1(0)=-[(1-4
\zeta^\1) \Rp^\1(0)/c_0]^{1/2}$ which can have a solution only for
$\zeta^\1 \leq \frac 14$.  Then the second derivative of Eq.
(\ref{fix.R.1}) implies
\begin{eqnarray}
  \p^3 \Rp^\1 (0^\pm) = \pm \sqrt{(1 - 2\zeta^\1)
    [- \p^2 \Rp^\1(0) /2 c_0]}.
  \label{jump}
\end{eqnarray}
The sign in front of the root is determined by the requirement that the
amplitude of $\p^2 \Rp^\1(\phi)$ should have a maximum at $\phi=0$.
Thus $\p^3 \Rp^\1 (0^\pm) \gtrless 0$, i.e. $\Rp^\1(\phi)$ is
discontinuous at $\phi=0$.  Consequently, $\p^4 \Rp^\1(\phi)$ contains
a singular contribution proportional to $\delta(\phi)$.

Now we look back to verify that this nonanalyticity does not
invalidate the RG analysis performed so far.  Since at the fixed point
$\Smm^\2$ is proportional to $\p \Rp^\1(\phi)$, discontinuities appear
if $\Smm^\2$ is derived twice with respect to one field argument. In
the derivation of Eq. (\ref{flow.Rp.o2}) some terms [for example
$\int_\bx \p_2^2 \Smm(\phi_1,0;\bx)$] have been dropped because of the
symmetry properties (\ref{symm.So.2}) of $\Smm$.  This amounts to the
implicit assumption $\p^3 \Rp(0)=0$, which seems arbitrary because of
the discontinuity of $\p^3 \Rp$.  However, in general there are
additional odd factors since the energy {\em functional} has to be
even in $\phi$. In the aforementioned example $\int_\bx \p_2^2
\Smm^\2(\phi_1,0;\bx) \propto \p \Rp^\1 (\phi_1) \p^3 \Rp^\1(0)$, i.e.
there is an additional odd factor $\p \Rp^\1 (\phi_1)$.  The
corresponding contribution to the functional (\ref{H.2}) then vanishes
after summing over the replica indices irrespective of the value of
$\p^3 \Rp(0)$.  Therefore there is no need to retain this term.

The situation becomes more severe as soon as a fourth derivative $\p^4
\Rp(\phi)$ appears.  Since the fixed point function $\Rpp^\2$ involves
second derivatives of $\Rp^\1$, this can happen where second
derivatives of $\Rpp^\2$ with respect to one field argument appear.
In the above flow equations (\ref{flow.fun}) and the resulting
fixed-point equations in order $\epsilon^2$ this is not the case.
However, in these equations we have neglected terms proportional to
temperature $T$. In particular, we have neglected contributions to the
flow of $\Rpp^\2$, which are proportional to $T \p_1^2
\Rpp^\2(\phi_1,\phi_2;\bx_{12}) \propto T \p^4 \Rp^\1(\phi)$.  From a
similar contribution to the flow equation of $\p^2 \Rp^\2$ one expects
$\p^4 \Rp^\1(0) \propto T^{-1}$, where $T$ is the effective
temperature that decreases on large scales.  Thus these particular
terms are proportional to $T^0$.  This means that temperature is a
dangerously irrelevant variable. (For a discussion of the role of
temperature in $d=1$ see Refs. \CITE{FisherDS+91:dp,Hwa+94:prb}.)
However, since $\Rpp^\2$ does not feed back into the fixed-point
equation for $\Rp^\1$, the roughness exponent is not affected in
leading order.  But these terms give contributions to the fluctuations
of free energy (that scale for large system sizes $L$ proportional to
$L^\theta$) which do not vanish in the limit $T \to 0^+$.

To sum up, we have shown universality of the roughness exponent to
leading order in $\epsilon$ with the SCRG method elaborated here.
Avoiding locality truncations, we have obtained the nonlocal
functional form of the fixed-point Hamiltonian. In particular, we have
determined higher cumulants of the effective pinning energy
distribution on large scales the implications of which will be
examined elsewhere.\cite{Scheidl00:un} 

Although in $d=2$ and $d=3$ the agreement between the RG result for
$\zeta$ and numerical values is quite good, it is of fundamental
interest to examine the possibility to extend the analytic theory
beyond this leading order.  It was argued by
Fisher\cite{FisherDS86:dr} and Balents and Fisher\cite{Balents+93}
that because of the nonanalyticity of the fixed point function
$\Rp(\phi)$ in $\cO(\epsilon)$ the next higher order after
$\cO(\epsilon^2)$ to the fixed-point equations should be
$\cO(\epsilon^{5/2})$.  Consequently, the subleading contribution to
$\zeta$ would be of order $\epsilon^{3/2}$.  However, their reasoning
is based on a $T=0$ argument.  As shown above, temperature is {\em
  not} a truly irrelevant variable and the validity of a $T=0$
argument for $T>0$ is questionable.  From preliminary
studies\cite{Dincer99,Scheidl00:un} we expect that the subleading
contribution to $\zeta$ is of order $\epsilon^2$.  The SCRG scheme
presented here lays the foundation for an extended analysis of the
functional RG beyond leading order in a systematic way.  In addition,
this method is free of the pathologies hampering the HCRG scheme and
better tractable than other SCRG schemes.


\section*{Acknowledgments}

The authors thank S. Bogner, T. Emig, T. Nattermann and H. Rieger for
helpful discussions.  S.S.  is indebted to H. Wagner for stimulating
suggestions already quite a long time ago.  This work was supported
financially by Deutsche Forschungsgemeinschaft through SFB 341.


\appendix
\section{Mode integration}
\label{app.int}

In this appendix we sketch the derivation of Eq. (\ref{zustand1}) from
Eq. (\ref{intro_Z}) and deduce the RG generator (\ref{gen.int})
following Polchinski.\cite{Polchinski84}

Starting from $G_\Lambda=G_<+G_>$, elementary manipulations lead to the
identity
\begin{equation}
  G_\Lambda^{-1} = G_>^{-1} - {G_>^{-1}(G_<^{-1}+G_>^{-1})}^{-1}
  G_>^{-1} .
  \label{split.G}
\end{equation}
We consider the partition sum of the pure interface as functional of
the propagator $G$,
\begin{eqnarray}
  \cZe[G]:= {\rm Tr}_\phi \ e^{-\frac{1}{2T}(\phi,G_\Lambda^{-1}\phi)}.
\end{eqnarray}
From $G_\Lambda=G_<+G_>$ one can immediately derive
\begin{eqnarray}
  \cZe[G_\Lambda]=\frac {\cZe[G_<] \cZe[G_>]}
  {\cZe[{(G_<^{-1}+G_>^{-1})}^{-1}]}.
  \label{factor.Zf}
\end{eqnarray}

Now the partition sum (\ref{intro_Z}) is transformed by the following
steps: we use Eq.  (\ref{split.G}), introduce the additional field
$\phi^<$, regroup fields in the bilinear Hamiltonian, use
(\ref{factor.Zf}), and substitute $\phi=\phi^<+\phi^>$: 
\beginwideapp
\begin{eqnarray}
  {\cal Z} 
  &=&
  {\rm Tr}_\phi \ e^{-\frac{1}{2T}(\phi,G_\Lambda^{-1}\phi)-
    \frac{1}{T} \cHp[\phi]} \nonumber \\
  &=&
  {\rm Tr}_\phi \ e^{-\frac{1}{2T}(\phi,G^{-1}_>\phi)+
    \frac{1}{2T}(G_>^{-1}\phi,
    (G^{-1}_<+G^{-1}_>){}^{-1} G^{-1}_>\phi)-
    \frac{1}{T} \cHp[\phi]}  \nonumber \\
  &=&
  \frac{1}{\cZe[{(G_<^{-1}+G^{-1}_>)}^{-1}]} \,
  {\rm Tr}_{\phi^<}{\rm Tr}_{\phi} \ 
  e^{-\frac{1}{2T}(\phi^<,(G^{-1}_<+G_> {}^{-1})\phi^<)+
    \frac{1}{T}
    (\phi^<,G_>^{-1}\phi)-\frac{1}{2T}(\phi,G_> {}^{-1}\phi)-
    \frac{1}{T} \cHp[\phi] } \nonumber \\
  &=&
  \frac {\cZe[G_\Lambda]}{\cZe[G_<] \cZe[G_>]}
  {\rm Tr}_{\phi^<}{\rm Tr}_{\phi} \
  e^{-\frac{1}{2T}(\phi^<,G^{-1}_< \phi^<)
    -\frac{1}{2T}((\phi-\phi^<),G^{-1}_>(\phi-\phi^<))-\frac{1}{T} 
    \cHp[\phi]}
  \nonumber \\
  &=&
  \frac {\cZe[G_\Lambda]}{\cZe[G_<] \cZe[G_>]}
  {\rm Tr}_{\phi^<}{\rm Tr}_{\phi^>}\
  e^{-\frac{1}{2T}(\phi^<,G^{-1}_< \phi^<)
    -\frac{1}{2T}(\phi^>,G^{-1}_> \phi^>)-\frac{1}{T} 
    \cHp[\phi^<+\phi^>]}
  \nonumber \\
  &=:&
  \frac {\cZe[G_\Lambda]}{\cZe[G_<]}
  {\rm Tr}_{\phi^<} \ e^{-\frac{1}{2T}(\phi^<,G^{-1}_< \phi^<)}
  \langle e^{-\frac{1}{T} \cHp[\phi^<+\phi^>]} \rangle_>
  \label{Z.trafo}
\end{eqnarray}
\endwideapp
This is Eq. (\ref{zustand1}) apart from the field independent
factors $\cZe$.

The RG generator (\ref{gen.int}) follows from evaluating the
expectation value in the last expression of Eq. (\ref{Z.trafo}). The
coarse-grained pinning Hamiltonian $\cHp'[\phi^<]$ for the slow
modes is defined by
\begin{eqnarray}
  e^{-\frac{1}{T} \cHp'[\phi^<]}:=
  \langle e^{-\frac{1}{T} \cHp[\phi^<+\phi^>]} \rangle_>
\end{eqnarray}
and is found from a cumulant expansion:
\begin{eqnarray}
  \cHp'[\phi^<] &=&
  \langle \cHp[\phi^<+\phi^>] \rangle_>
  \nonumber \\ &&
  - \frac 1{2T} \langle \cHp^2[\phi^<+\phi^>] \rangle_>^\cum
  + \cdots
  \label{H.primed}
\end{eqnarray}
A Taylor expansion of $\cHp[\phi^<+\phi^>]$ in $\phi^>$ yields
($\phi_i^> := \phi^>(\bx_i)$ etc.)
\begin{eqnarray*}
  \langle \cHp[\phi^<+\phi^>] \rangle_> &=& \cHp[\phi^<]
  \nonumber \\
  && + \frac{T}{2} \int_{12} \frac{\delta^2 \cHp}{\delta
    \phi_1^> \delta \phi_2^>} G^>_{12} 
  + {\cal O}(G^2_>) \,,
  \\
  \langle \cHp^2[\phi^<+\phi^>]  \rangle_>^\cum&=& 
  T \int_{12} 
  \frac{\delta\cHp} {\delta \phi^>_1} 
  \frac{\delta\cHp} {\delta \phi^>_2} 
  G^>_{12} + {\cal O} (G^2_>) \,.
\end{eqnarray*}
All neglected terms arising from higher cumulants of higher orders of
Taylor expansion involve at least two propagators $G_>$. Since $G_>$
is bounded, terms of order $G_>^2$ result in functionals of order
$d\ell^2$ for small $d\ell$ and do not contribute to the differential
RG.  Plugging the contributions of the last equations into
(\ref{H.primed}) result in the generator (\ref{gen.int}). For further
details see also Ref.  \CITE{Zinn-Justin93}.


\section{Kernels, coefficients}
\label{app.kern}

Here we give explicit expressions for the kernel functions and the
coefficients entering the flow equations.  This is interesting for
illustrative purposes but also important to demonstrate, that they are
well defined.  We start from the Gaussian cutoff function
(\ref{f.Gauss}) and perform calculations in $d=4$.  We find from Eqs.
(\ref{intro.g}), (\ref{sol.sigma}), (\ref{def.c}), and (\ref{rho.sol})
\begin{eqnarray*}
  g(\bk) &=& \frac 1{\gamma \Lambda^2} e^{- k^2/2 \Lambda^2}
  \\
  \sigma(\bk) &=& \frac 1{\gamma k^2} \left( 1 - e^{- k^2/2 \Lambda^2}
  \right)
  \\
  c(\bk) &=& \frac {g_0}2 \sigma(\bk/\sqrt 2)
  \\
  \rho(\bk) &=& c_0 \int_0^1 dt \frac 1t \left\{ 1 - \frac {4
      \Lambda^2}{t^2 k^2} \left[ 1 - e^{-t^2 k^2/4 \Lambda^2} \right]
  \right\}
  \\
  &\to& \left\{
    \begin{array}{ll}
      \frac{c_0 k^2}{16 \Lambda^2} & \textrm{ for } k \to 0,
      \\
      c_0 \ln \frac{k}{2\Lambda} & \textrm{ for } k \to \infty.
    \end{array}
    \right.
\end{eqnarray*}
with $g_0 := g(\bx=\bN)= \frac {\Lambda^2}{ 4 \pi^2 \gamma}$ and $c_0=
\frac 1{(4 \pi \gamma)^2}$.  These kernels are apparently well-defined
in $d=4$ and analytic for $\bk \to \bN$.

\bibliographystyle{prsty}
\bibliography{library,frg}

\begin{thebibliography}{10}
\bibitem{note.scal}
At first sight it may not be obvious that the first term in Eq.
  (\ref{gen.scal}) accounts for the rescaling of lengths. This functional form
  for rescaling was used previously, e.g., in Ref. \CITE{Wegner76}, Eq.
  (3.158).
\end{thebibliography}

\end{multicols}
\end{document}